\documentclass[preprint2]{emulateapj}

\begin{document}

\title{Search for gamma-ray emission from four accreting millisecond
pulsars with \textit{Fermi}/LAT}

\author{Yi Xing and Zhongxiang Wang}
\affil{Shanghai Astronomical Observatory, Chinese Academy of Sciences,\\
80 Nandan Road, Shanghai 200030, China}

\begin{abstract}
We report our search for $\gamma$-ray emission in the energy range from 
100 MeV to 300 GeV from four Accreting Millisecond Pulsars (AMPs), 
SAX J1808.4$-$3658, IGR J00291$+$5934, XTE J1814$-$338, and XTE J0929$-$314. 
The data are from four-year observations carried out by Large Area Telescope 
(LAT) onboard the \textit{Fermi} $\gamma$-ray Space Telescope. 
The AMPs were not detected, and their $\gamma$-ray luminosity upper 
limits we obtain are $5.1\times10^{33}$ ergs s$^{-1}$ for 
SAX J1808.4$-$3658, $2.1\times10^{33}$ ergs s$^{-1}$ for IGR J00291$+$5934, 
$1.2\times10^{34}$ ergs s$^{-1}$ for XTE J1814$-$338, 
and $2.2\times10^{33}$ ergs s$^{-1}$ for XTE J0929$-$314. 
We compare our results with $\gamma$-ray irradiation luminosities required 
for producing optical modulations seen from the companions in the AMPs, 
which has been suggested by Takata et al. (2012), and our upper 
limits have excluded $\gamma$-ray emission as the heating source in these 
systems except XTE J0929$-$314, the upper limit of which is not deep enough. 
Our results also do not support the model proposed by Takata et al. (2012) that
relatively strong $\gamma$-ray emission could arise from the outer gap of 
a high-mass neutron star controlled by the photon-photon pair-creation 
for the AMPs.
Two AMPs, SAX J1808.4$-$3658 and IGR J00291$+$5934, have the measurements of
their spin-down rates, and we derive the upper limits of their $\gamma$-ray 
conversion efficiencies, which are 57\% and 3\%, respectively. We discuss the
implications to the AMP systems by comparing the efficiency upper limit values
with that of 20 $\gamma$-ray millisecond pulsars (MSP) detected 
by \textit{Fermi} and the newly discovered transitional MSP binary J1023+0038.
\end{abstract}


\section{Introduction}
Accreting millisecond pulsars (AMPs) are all in low mass X-ray binary (LMXB)
systems, spinning at frequencies more than 100 Hz and containing companion
stars with masses less than $\mathrm{1\, M_{\odot}}$ \citep{pw12}.
The periodical pulsed emission from AMPs is accretion powered, which
distinguishes them from normal radio millisecond pulsars (MSPs) that are powered
by rotation. AMPs are believed to be the progenitors of MSPs spun up from
old slowly rotating neutron stars (NSs) by the recycling
scenario \citep{alp+82, bv91, sri10}, that is when a NS in
a binary system switches off the pulsed radio emission, the mass and angular
momentum of the companion star are transferred to the NS through
accretion, producing a recycled pulsar that eventually spins at a period of
milliseconds. The currently known AMPs are all transient systems with recurrent times of
years. Evolution of their accretion disks leads to weeks-long outbursts,
during which pulsed X-ray emission is detected.

The first known AMP SAX J1808.4$-$3658 was found by the \textit{BeppoSAX}
mission in 1996 \citep{int+98} and discovered to have pulsed X-ray emission
with the \textit{Rossi X-ray Timing Explorer} (RXTE) during its 1998
outburst \citep{wv98, cm98}. Since then, over a dozen of AMP systems
have been discovered \citep{pw12}. One interesting property of AMP binaries 
was first pointed out by \citet{bur+03}: because the orbital optical 
modulation seen in SAX J1808.4$-$3658
has a larger than expected amplitude in the quiescent state \citep{hom+01},
the AMP probably switches to be rotation-powered and thus is able to provide
the required energy output to heating of the companion.
Followup observations of SAX J1808.4$-$3658 at X-ray and optical energies have
firmly verified the inconsistency between the large optical modulation and
low X-ray luminosity in quiescence \citep{cam+04, del+08, wan+09}.
X-ray emission from accreting compact stars normally is the heating source
to cause orbital optical modulations seen in LMXBs (e.g., \citealt{vm95}).
In addition, the inconsistency has also been seen in two other AMP systems,
IGR~J00291$+$5934 and XTE~J1814$-$338 \citep{dav+07,jts08,dav+09}, which 
suggests a common property of being able to switch to be radio pulsars 
in quiescence for AMP systems, although no direct evidence has been found
from radio observations of them \citep{burgay+03}.

The more recent discovery of the millisecond pulsar binary PSR J1023$+$0038
\citep{arc+09} has strengthened the likelihood of the radio-pulsar
switching property of the AMP systems. The binary is considered to be 
the first system found at the end of its evolution from a LMXB to a MSP binary,
as a short-term accretion disk was seen in it \citep{wat+09, wwm12}.
$\gamma$-ray emission from PSR J1023$+$0038 was also found from observations 
carried out by
\textit{Fermi} Gamma-ray Space Telescope \citep{tam+10}. Considering 
PSR~J1023$+$0038 as an end product of LMXB evolution closely related to
AMP systems, \citet{tct12} studied possible $\gamma$-ray emission
mechanisms for the AMP systems known with large-amplitude optical modulations 
(see Table 1). In their work, they attributed the optical modulations of 
the AMP systems to $\gamma$-ray irradiation of the companions, which 
required $\gamma$-ray luminosity 
$L_{\gamma}$ $>$ $10^{32}-10^{34}$ ergs~s$^{-1}$. 
The predicted $\gamma$-ray emission is
possibly detectable by \textit{Fermi}, and any detection would provide strong
evidence for the existence of radio pulsars in these AMP systems in quiescence.
We thus searched for $\gamma$-ray emission from the four AMPs in the data
obtained by \textit{Fermi}  
and report our results in this paper. The general properties of the four AMPs 
are summarized in Table 1.

In \S~\ref{sec:obs} we describe the data from the
\textit{Fermi} Large Area Telescope (LAT) observations and our  
analysis of the data for the four AMPs. We provide the results in 
\S~\ref{sec:res} and discuss the implications of our results in 
\S~\ref{sec:dis}.

\section{Observation and data analysis}
\label{sec:obs}

LAT is an imaging instrument onboard the \textit{Fermi} Gamma-ray Space 
Telescope. Its main 
objective is to conduct continuous $\gamma$-ray monitoring of a large sample 
of sources in energy range from 20 MeV to 300 GeV with much improved 
sensitivity compared with former $\gamma$-ray telescopes \citep{atw+2009}. 
In our data analysis we selected LAT events within 15 degrees centered at 
the positions of the four AMPs during time period from the start of 
the releasing of \textit{Fermi} public data 2008-08-04 15:43:36 (UTC) to 
2012-07-08 
18:59:57 (UTC) from the \textit{Fermi} Pass 7 database. We included events 
in an energy range from 100 MeV to 300 GeV. Following recommendations of 
the LAT team, we required events zenith angle to be less than 100 degrees 
to prevent the Earth's limb contamination, and excluded events in 
time intervals when spacecraft events affected the quality of the LAT data.

\subsection{Maximum likelihood analysis}

We rejected events below 200 MeV and included only events in energy range 
from 200 MeV to 300 GeV for the likelihood analysis, because the instrument 
response function of LAT has relative large uncertainties in lower energy 
range. We included all sources within 20 degrees centered at the four AMPs 
to make the source models based on the \textit{Fermi} 2-year catalog. 
Most of them are point sources, while a few of them are diffuse sources. 
The spectral function forms of these sources were selected the same as those 
in the catalog. We let the spectral normalization parameters of sources 
within 3 degrees from the four AMPs free, and fixed all the other parameters 
of sources included in our source models to their catalog values. The four 
AMPs we are interested in were not included in the Fermi catalog. We modeled 
the $\gamma$-ray spectra of them with a exponentially cut-off power law which 
is the typical spectral form of pulsars \citep{aaa+09, aaa+10}. 
The exponentially cut-off power law can be expressed by 
$dN/dE=N_{0}E^{-\Gamma}exp[-(E/E_{cut})]$, 
where $\Gamma$ is the spectral index and 
$E_{cut}$ is the cut-off energy. In addition we considered 
the Galactic and extragalactic diffuse emission using the model 
gal\_2yearp7v6\_v0.fits and the spectrum file iso\_p7v6source.txt, 
respectively. The value of the Galactic diffuse emission model and 
the normalization of the extragalactic diffuse emission model were let free.

We performed standard binned likelihood analysis with the LAT science tools 
software package v9r27p1. The binned likelihood used events in a square 
region instead of a circle, so we selected events inside 
a $\mathrm{20^{o}\times20^{o}}$ region centered at each interested source, 
which is nearly the inscribed square of the circular region we selected above. 
We obtained the Test Statistic (TS) of the four AMPs, and the TS maps of
$\mathrm{3^{o}\times3^{o}}$ regions centered at each of the four AMPs are
shown in Figure~1. Each of the maps was
extracted by assuming a putative source and performing binned likelihood 
analysis to obtain the TS value at each grid point of the regions. 
The sources in the source models outside of the regions, and the Galactic 
and extragalactic diffuse emission were considered. 
We found that IGR J00291$+$5934, XTE J1814$-$338, and XTE J0929$-$314 had 
TS values of $\sim$ 0, and SAX J1808.4$-$3658 had a TS value of $\sim$ 3. 
A TS value is 
approxmiately the square of the detection significance of a source and it 
can be used to assess whether a source exists or not. Actually in the 
\textit{Fermi} source catalog only sources with TS values greater than 25 were 
included, which corresponds to the detection significance of 
4.6$\sigma$ \citep{aaa+2010}. The TS values of the four AMPs thus indicate 
that they were not detected by \textit{Fermi}/LAT in 200 MeV $-$ 300 GeV band.

\subsection{Flux Upper limit Calculation}
The $\gamma$-ray flux upper limits of these four AMPs were estimated by 
the binned likelihood analysis with the source models we described above. 
We fixed the $\Gamma$ and $E_{cut}$ parameters of 
the spectral models of the four AMPs to 1.4 and 1.6 GeV respectively, 
which are the averaged spectral parameters of the eight MSPs \textit{Fermi} 
detected during the first six months monitoring \citep{aaa+09, aaa+10} 
and derived from the \textit{Fermi} 2-year catalog (see Table 2). The spectral 
normalization factors were let free. We extracted the upper limits of 
normalization factors by increasing these factors until the maximum likelihood 
values decreased by e/2 in logarithm, following the procedure introduced 
by the LAT team. Then using the upper limits of normalization factors we 
derived the 95$\%$ flux upper limits in energy range from 100 MeV to 300 GeV 
by integrating the spectral models. Assuming isotropic $\gamma$-ray emission, 
the upper limits of the $\gamma$-ray luminosities of these four sources were 
calculated, where the distance values of the sources listed in Table 1 were 
used.

\subsection{Timing analysis}
We performed timing analysis of the \textit{Fermi}/LAT observations of 
the four AMPs to search for any $\gamma$-ray pulsations. We included events 
in energy range from 100 MeV to 300 GeV within 1 degree centered at 
the positions of these four sources, and folded them into 20 phase bins 
using X-ray ephemerides (Table 1). 
The frequency derivatives were considered for the sources with the measurements
of their spin-down rates. We obtained the folded light curves of these 
sources (Figure 2) and made $\chi^{2}$ test and H test to identify possible 
presence of $\gamma$-ray pulsations. The four folded light curves have 
$\chi^{2}$ values in the range of 11$-$28 which are comparable to 
the 19 degrees of freedom \citep{lar96}, and H values in the range of 
0.4$-$10.8 which are significantly smaller than the value (H$=$42) 
used by the LAT team to comfirm $\gamma$-ray pulsations \citep{aaa+10}, 
thereby indicating no significant $\gamma$-ray pulsations from these four AMPs 
were detected. In addition, since our targets are all located in the Galactic
plane, contamination from diffuse emission 
and nearby sources may prevent the detections of $\gamma$-ray pulsations. 
We thus tested to decrease the radii of the target regions from 1 degree 
to 0.5 degree to search for $\gamma$-ray pulsations. However no 
significant $\gamma$-ray pulsations were detected either with $\chi^{2}$ values of 7$-$32 (19 degrees of freedom) and H values of 0.1$-$9.3. 

For SAX J1808.4$-$3658 and IGR J00291$+$5934, which had X-ray outbursts 
after August 2008 under the \textit{Fermi} coverage, we also searched their $\gamma$-ray pulsations during the outbursts. SAX J1808.4$-$3658 had the X-ray outbursts in 2008 starting at MJD 54730 with $\nu$ of $\mathrm{400.97521009~Hz}$ and $\dot{\nu}$ of $0.6\times10^{-13}$ Hz s$^{-1}$ \citep{hpc+09}, and in 2011 starting at MJD 55865 with $\nu$ of $\mathrm{400.97520981~Hz}$ and non-significant $\dot{\nu}$ of $<4\times10^{-13}$ Hz s$^{-1}$ \citep{pbg+12}. IGR J00291$+$5934 had two X-ray outbursts in 2008 starting at MJD 54691 and MJD 54727 \citep{lew10}. X-ray observations in MJD 54691.94--54696.77 revealed $\nu$ of $\mathrm{598.89213061~Hz}$ and $\dot{\nu}$ of $12.3\times10^{-13}$ Hz s$^{-1}$, and in MJD 54730.51--54742.89 revealed $\nu$ of $\mathrm{598.89213046~Hz}$ and $\dot{\nu}$ of $5.7\times10^{-13}$ Hz s$^{-1}$ \citep{p10}. We searched $\gamma$-ray pulsations of SAX J1808.4$-$3658 during MJD 54735--54760 and MJD 55870--55886, and of IGR J00291$+$5934 during the X-ray observations of its two outbursts, using their spin parameters given above. No detections of $\gamma$-ray pulsations were found from these two sources.

\subsection{Analysis of the Data in the Quiescent State}

The data used in our analyses in \S~2.1 and \S~2.2 were nearly four years 
from 2008--2012.  As given
in \S~2.3, during the time period SAX J1808.4$-$3658 had two X-ray outbursts 
and IGR J00291$+$5934 had two close X-ray outbursts. According to the current models for the $\gamma$-ray
emission mechanism for pulsars (see discussion in \S~4), accretion of the MSPs in these 
two binaries in the outbursts should have quenched $\gamma$-ray emission 
from their magnetospheres.
Although the durations of the outbursts were short compared to the 4-year 
length of the total data, we checked whether our non-detections of the
two sources and upper limit calculations would be significantly affected.
The outburst state of the two sources 
lasted about 20--50 days \citep{hpc+09,pbg+12,lew10}, and optical observations 
showed the brightening of the accretion disks started several tens of days 
before the outbursts \citep{wan+13,lew10}. 
We excluded \textit{Fermi}/LAT data from 10 days prior to and 50 days 
after the starting of each outburst and performed likelihood analysis using the data 
only in the quiescent state given in Table 1. With the same event selections and source 
models, the TS values we obtained in quiescent state are $\sim$ 3 for SAX J1808.4$-$3658 
and $\sim$ 0 for IGR J00291$+$5934, and the $\gamma$-ray luminosity upper limits 
are $5.0\times10^{33}$ ergs s$^{-1}$ for SAX J1808.4$-$3658 and $2.3\times10^{33}$ ergs s$^{-1}$ 
for IGR J00291$+$5934, which are nearly the same as that we obtained using the total data sets.

\section{Results}
\label{sec:res}

We found that the four AMPs included in our work were not detected by 
\textit{Fermi}/LAT at $\gamma$-ray energies. We obtained upper limits of 
their $\gamma$-ray luminosities, that are 
$5.1d^{2}_{3.5}\times10^{33}$ ergs s$^{-1}$ for SAX J1808.4$-$3658, 
$2.1d^{2}_{5}\times10^{33}$ ergs s$^{-1}$ for IGR J00291$+$5934, 
$1.2d^{2}_{8}\times10^{34}$ ergs s$^{-1}$ for XTE J1814$-$338, 
and $2.2d^{2}_{6}\times10^{33}$ ergs s$^{-1}$ for XTE J0929$-$314, 
where $d$ is the source distance with its subscript indicating the value 
(in units of kpc; Table 1) used. These results are summarized in Table 1. 
Two AMPs---SAX J1808.4$-$3658 and IGR J00291$+$5934---have measurements 
of their spin-down rates (Table 1), through which we estimated the spin-down 
luminosities by $\dot{E}=4\pi^{2}I\nu\dot{\nu}$, where $I$ is the 
NS moment of inertia and assumed to have a canonical value of 
$I=10^{45}$~g~cm$^{2}$. We calculated the upper limit of $\gamma$-ray 
conversion efficiency $\eta_{\gamma}=L_{\gamma}/\dot{E}$ for these 
two sources with the upper limits of $\gamma$-ray luminosities we obtained 
above. SAX J1808.4$-$3658 has $\eta_{\gamma}$ of $<$ 0.57$d^{2}_{3.5}$,
and IGR J00291$+$5934 has $\eta_{\gamma}$ of $<$ 0.03$d^{2}_{5}$. 
The latter provides a strong constraint on the $\gamma$-ray emission from 
this source.

\section{Discussion}
\label{sec:dis}

We searched $\gamma$-ray emission from four AMPs, SAX J1808.4$-$3658, 
IGR J00291$+$5934, XTE J1814$-$338, and XTE J0929$-$314 with 
\textit{Fermi}/LAT observations. The four sources were not detected by 
\textit{Fermi}/LAT in the 200 MeV to 300 GeV energy range. We obtained 
the 100 MeV to 300 GeV $\gamma$-ray luminosity upper limits of these 
four sources. In addition for two of them, SAX J1808.4$-$3658 and 
IGR J00291$+$5934, with the measurements of the spin-down rates,
we estimated the 100 MeV to 300 GeV $\gamma$-ray conversion efficiency upper 
limits. The obtained values of these sources can be used to compare with 
model predictions of AMPs and conversion efficiencies of $\gamma$-ray MSPs, 
which provides constraints on their properties including $\gamma$-ray emission 
mechanisms.

Depending on different assumptions, a wide range of $L_{\gamma}$ values 
for the four AMPs have been predicted by \citet{tct12}. Futhermore they have 
suggested that optical modulations seen in the AMP binaries could be 
caused by $\gamma$-ray irradiation of the companion stars. 
The irradiation $\gamma$-ray luminosities of these sources required 
for optical modulations were used to constrain the theoretical emission 
models. For SAX J1808.4$-$3658, IGR J00291$+$5934 and XTE J1814$-$338, 
the inferred $\gamma$-ray irradiation luminosities of 
$>$ $10^{34}$ ergs s$^{-1}$ favored the outer gap model with a high-mass 
NS controlled by the photon-photon pair-creation between $\gamma$-rays 
and X-rays from full surface cooling emission, or the outer gap model 
controlled by the magnetic pair-creation process \citep{tct12}. Such a high 
luminosity can not be produced by the outer gap model controlled by 
the photon-photon pair-creation between $\gamma$-rays and X-rays from 
the heated polar cap because an unreasonably high NS magnetic field is 
required. For XTE J0929$-$314 with a lower inferred $\gamma$-ray irradiation 
luminosity of $10^{32}-10^{33}$ ergs s$^{-1}$, the outer gap model with a 
low-mass NS controlled by the photon-photon pair-creation between 
$\gamma$-rays and X-rays from full surface cooling emission might be prefered \citep{tct12}.

In our work, we found that the $\gamma$-ray luminosities of 
SAX J1808.4$-$3658 and IGR J00291$+$5934 observed by \textit{Fermi}/LAT 
had the upper limits of $5.1d^{2}_{3.5}\times10^{33}$ ergs s$^{-1}$ 
and $2.1d^{2}_{5}\times10^{33}$ ergs s$^{-1}$ respectively, 
which are significantly lower than the required $\gamma$-ray irradiation 
luminosities of these two sources. For XTE J1814$-$338 and XTE J0929$-$314, 
the $\gamma$-ray luminosity upper limits 
of $1.2d^{2}_{8}\times10^{34}$ ergs s$^{-1}$ and 
$2.2d^{2}_{6}\times10^{33}$ ergs s$^{-1}$ are consistent with 
the required $\gamma$-ray irradiation luminosities. However for 
XTE J1814$-$338, the $\gamma$-ray irradiation luminosity lower limit given
by \citet{tct12} is nearly equal to the luminosity upper limit we obtained. 
Considering the actual luminosities of these AMPs might be significantly 
smaller than the upper limits, the \textit{Fermi} observations of 
these sources indicate that the $\gamma$-ray emission of three 
AMPs, SAX J1808.4$-$3658, IGR J00291$+$5934, and XTE J1814$-$338, are 
lower than the values needed to interpret the optical modulations. Thus 
$\gamma$-ray irradiation from the AMPs should not be the main heating 
source of their companions. The irradiation from release of the rotational 
energy of the pulsars should be prefered, which indeed provide the required 
energy output of $10^{33}-10^{34}$ ergs s$^{-1}$ to heat 
the companions (\citealt{bur+03}, Table 1), although there are a 
few unresolved theoretical uncertainties in this model \citep{bur+03,tct12}.

$\gamma$-ray emission from pulsars is generally agreed to originate 
from accelerations of charged particles, while it is still uncertain 
where the accelerations occur. At present there are polar cap, outer gap, and
slot gap models to be advanced to interpret the high energy emission of 
pulsars. The polar cap model \citep{rs75} has acceleration regions near 
the surface of a NS, which predicts pulsed emission aligned with 
the magnetic poles. The model fails to interpret the $\gamma$-ray pulsed 
profiles of pulsars because most of pulsars have two sharp peaks which are 
not aligned with radio peaks \citep{aaa+09,aaa+10}. The outer gap \citep{mh04} 
and slot gap models \citep{chr86} have acceleration regions along 
the last open field lines to near the light cylinder, starting from 
the null surface and the polar cap, respectively. The two models both predict to
have wide-fan beams which are not aligned with magnetic poles so that 
they are more prefered to interpret $\gamma$-ray emission from pulsars. 
Comparing with that on the basis of the outer gap model predicted 
by \citet{tct12}, our results basically rule out
their model of a high-mass NS controlled by the photon-photon 
pair-creation between $\gamma$-rays and X-rays from full surface cooling 
emission for SAX J1808.4$-$3658, IGR J00291$+$5934 and XTE J1814$-$338. 
For XTE J0929$-$314, the outer gap emission with a low-mass NS is predicted and
our results is still consistent with their model prediction. 

The newly born MSP PSR J1023$+$0038 from an LMXB system has been 
detected by \textit{Fermi}/LAT with a detection significance 
of $\sim$ 7$\sigma$ in the 200 MeV to 20 GeV band \citep{tam+10}, which 
greatly motivated our search for $\gamma$-ray emission from the AMPs to 
verify the possible existence of rotation-powered pulsars in the quiescent 
state. However, although many $\gamma$-ray MSPs have been detected by 
\textit{Fermi}/LAT \citep{aaa+09,aaa+10}, no $\gamma$-ray AMPs 
have been found. In our work we gave the upper limits of 100 MeV to 300 GeV 
$\eta_{\gamma}$ for two AMPs, that are 57$\%$ for SAX J1808.4$-$3658 
and 3$\%$ for IGR J00291$+$5934. We compared them with MSPs detected by 
\textit{Fermi}/LAT. In the \textit{Fermi} 2-year catalog there are 38 
possible $\gamma$-ray MSPs, among which 20 are confirmed MSPs that 
have typical spectral form of pulsars (exponentially cut-off power law) and 
pulsations detected in $\gamma$-ray band. A summary of these 20 MSPs is given
in Table 2. We calculated 
the 100 MeV to 100 GeV $\eta_{\gamma}$ of these MSPs with 
flux values given in the catalog and timing parameters listed in Table 2. 
These sources' $\eta_{\gamma}$ are plotted in Figure 3, with 
uncertainties derived from flux uncertainties given in the catalog. 
In addition the uncertainties related to distance measurements are 
approximately 60$\%$ presuming 30$\%$ distance uncertainties. 
As can be seen, the $\gamma$-ray conversion efficiency 
of $<$ 3$\%$ of IGR J00291$+$5934 lies in the very lower range of that 
of $\gamma$-ray MSPs. This upper limit is a strict constraint on 
the $\gamma$-ray emission of IGR J00291$+$5934, which likely suggests that this 
source may not emit in $\gamma$-ray band, or its $\gamma$-ray beam, at least 
the bright portion of its $\gamma$-ray beam, may not cross the earth. 
For SAX J1808.4$-$3658, its $\gamma$-ray conversion efficiency is $<$57$\%$, 
which is not sufficiently low suggesting this source as a candidate
$\gamma$-ray source for further monitoring with longer observation time.

It can be noted that PSR J1023$+$0038 has been detected with the lowest 
conversion efficiency of $\sim$ 0.5$\%$ in $>$100 MeV band (Figure 3; 
derived from spectral parameters given in \citealt{tam+10}). This may 
suggest that $\gamma$-ray emission of MSPs is related to their evolutionary 
stage. The fast rotating AMPs may have almost no $\gamma$-ray emission, but 
at the end of the accreting phase the newly born MSPs begin to emit 
in $\gamma$-ray band with low $\gamma$-ray conversion efficiencies.
We have also searched for pulsed emission from J1023+0038 in the 
\textit{Fermi} data from 2008 to the present time (data in \citealt{tam+10}
were to 2010 July), but have not found any detection. Comparisons of the properties
of its $\gamma$-ray pulsed emission with that of its radio emission may have 
implications to what are expected from AMPs. In addition, we note that 
no $\gamma$-ray MSPs are located in the low-right corner of Figure 3. We may 
approximately define a \textquoteleft death line\textquoteright, 
$\eta_{\gamma}=10^{0.32\tau_9-4.3}$ 
where $\tau_9$ is the characteristic age in units of 10$^9$ yrs, by making 
all $\gamma$-ray MSPs above the line. 
Comparing to a theoretical death line, for example $\lg\dot{P}=-14.1+2.83\lg P$, which was recently derived by \citet{wang11} for $\gamma$-ray MSPs on the basis of the outer gap model, the line is nearly straight in Figure 3 if a constant $L_{\gamma}$ is considered (dotted line; $L_{\gamma}\approx2\times10^{30}$ ergs s$^{-1}$ is used to draw the line). \citet{tct12} found that $L_{\gamma}\sim10^{34}\tau_{9}^{-5/4}$ ergs s$^{-1}$ fits luminosities of $\gamma$-ray MSPs relatively well [derived from the outer gap model controlled by magnetic pair-creation; cf. Eq. (32)], which implies that $\eta_{\gamma}$ is generally inversely proportional to $\tau$ [note $\dot{E}\sim\dot{P}/P^{3}\sim1/(\tau P^{2})$ or $\dot{E}\sim P^{-0.17}$ when the death line of $\lg\dot{P}=-14.1+2.83\lg P$ is used]. The line we define instead suggests that older detectable MSPs tend to have higher $\eta_{\gamma}$. 
If this is a true feature, SAX J1808.4$-$3658 would have no $\gamma$-ray 
emission detectable because its conversion efficiency upper limit is very 
close to the line. Hopefully with more MSPs detected by 
\textit{Fermi} in the near future, whether this line exists 
or not can be verified.
 
Finally the complex regions near the Galactic plane, in which 
the four AMPs are located, certainly affects the $\gamma$-ray detections 
and the upper limit evaluations of them. As can be seen from 
the TS maps (Figure 1), while the Galactic diffuse emission was removed 
from the source regions at a satisfactory level, there is excess in these 
regions in addition to those corresponding to the nearby $\gamma$-ray sources 
in the catalog, particularly for SAX J1808.4$-$3658. The TS excess near 
the location of SAX J1808.4$-$3658 is not sufficiently strong for 
the $\gamma$-ray detection of this source due to the low TS value. On
the other hand if we consider the TS excess as indications of 
other possible weak sources, the luminosity upper limit of the source should be slightly 
lower than that we have obtained.
It is also worth noting that our likelihood analysis of $\gamma$-ray emission 
from the four AMPs have 
uncertainties including that from the LAT instrument response function, 
Galactic diffuse emission, and nearby sources' emission. Data of longer 
monitoring by \textit{Fermi}/LAT will help reduce the uncertainties and 
thus provide tighter constraints on the emission properties of the AMPs.

\acknowledgments
We thank the anonymous referee for valuable suggestions. This research was supported by National Basic Research Program of China
(973 Project 2009CB824800) and National Natural Science
Foundation of China (11073042).
ZW is a Research Fellow of the 
One-Hundred-Talents project of Chinese Academy of Sciences.


\begin{thebibliography}{48}
\expandafter\ifx\csname natexlab\endcsname\relax\def\natexlab#1{#1}\fi

\bibitem[{{Abdo} {et~al.}(2009){Abdo}, {Ackermann}, {Ajello}, {Atwood},
  {Axelsson}, {Baldini}, {Ballet}, {Barbiellini}, {Baring}, {Bastieri},
  {Baughman}, {Bechtol}, {Bellazzini}, {Berenji}, {Bignami}, {Blandford},
  {Bloom}, {Bonamente}, {Borgland}, {Bregeon}, {Brez}, {Brigida}, {Bruel},
  {Burnett}, {Caliandro}, {Cameron}, {Camilo}, {Caraveo}, {Carlson},
  {Casandjian}, {Cecchi}, {{\c C}elik}, {Charles}, {Chekhtman}, {Cheung},
  {Chiang}, {Ciprini}, {Claus}, {Cognard}, {Cohen-Tanugi}, {Cominsky},
  {Conrad}, {Corbet}, {Cutini}, {Dermer}, {Desvignes}, {de Angelis}, {de Luca},
  {de Palma}, {Digel}, {Dormody}, {do Couto e Silva}, {Drell}, {Dubois},
  {Dumora}, {Edmonds}, {Farnier}, {Favuzzi}, {Fegan}, {Focke}, {Frailis},
  {Freire}, {Fukazawa}, {Funk}, {Fusco}, {Gargano}, {Gasparrini}, {Gehrels},
  {Germani}, {Giebels}, {Giglietto}, {Giordano}, {Glanzman}, {Godfrey},
  {Grenier}, {Grondin}, {Grove}, {Guillemot}, {Guiriec}, {Hanabata}, {Harding},
  {Hayashida}, {Hays}, {Hobbs}, {Hughes}, {J{\'o}hannesson}, {Johnson},
  {Johnson}, {Johnson}, {Johnson}, {Johnston}, {Kamae}, {Katagiri}, {Kataoka},
  {Kawai}, {Kerr}, {Kn{\"o}dlseder}, {Kocian}, {Kramer}, {Kuss}, {Lande},
  {Latronico}, {Lemoine-Goumard}, {Longo}, {Loparco}, {Lott}, {Lovellette},
  {Lubrano}, {Madejski}, {Makeev}, {Manchester}, {Marelli}, {Mazziotta},
  {McConville}, {McEnery}, {McLaughlin}, {Meurer}, {Michelson}, {Mitthumsiri},
  {Mizuno}, {Moiseev}, {Monte}, {Monzani}, {Morselli}, {Moskalenko}, {Murgia},
  {Nolan}, {Norris}, {Nuss}, {Ohsugi}, {Omodei}, {Orlando}, {Ormes}, {Paneque},
  {Panetta}, {Parent}, {Pelassa}, {Pepe}, {Pesce-Rollins}, {Piron}, {Porter},
  {Rain{\`o}}, {Rando}, {Ransom}, {Ray}, {Razzano}, {Rea}, {Reimer}, {Reimer},
  {Reposeur}, {Ritz}, {Rochester}, {Rodriguez}, {Romani}, {Roth}, {Ryde},
  {Sadrozinski}, {Sanchez}, {Sander}, {Saz Parkinson}, {Scargle}, {Schalk},
  {Sgr{\`o}}, {Siskind}, {Smith}, {Smith}, {Spandre}, {Spinelli}, {Stappers},
  {Starck}, {Striani}, {Strickman}, {Suson}, {Tajima}, {Takahashi}, {Tanaka},
  {Thayer}, {Thayer}, {Theureau}, {Thompson}, {Thorsett}, {Tibaldo}, {Torres},
  {Tosti}, {Tramacere}, {Uchiyama}, {Usher}, {Van Etten}, {Vasileiou},
  {Venter}, {Vilchez}, {Vitale}, {Waite}, {Wallace}, {Wang}, {Watters}, {Webb},
  {Weltevrede}, {Winer}, {Wood}, {Ylinen}, \& {Ziegler}}]{aaa+09}
{Abdo}, A.~A., {Ackermann}, M., {Ajello}, M., {et~al.} 2009, Science, 325, 848

\bibitem[{{Abdo} {et~al.}(2010{\natexlab{a}}){Abdo}, {Ackermann}, {Ajello},
  {Allafort}, {Baldini}, {Ballet}, {Barbiellini}, {Bastieri}, {Bechtol},
  {Bellazzini}, {Berenji}, {Blandford}, {Bloom}, {Bonamente}, {Borgland},
  {Bouvier}, {Bregeon}, {Brez}, {Brigida}, {Bruel}, {Burnett}, {Buson},
  {Caliandro}, {Cameron}, {Camilo}, {Caraveo}, {Carrigan}, {Casandjian},
  {Cecchi}, {{\c C}elik}, {Chekhtman}, {Cheung}, {Chiang}, {Ciprini}, {Claus},
  {Cognard}, {Cohen-Tanugi}, {Conrad}, {Corbet}, {DeCesar}, {Dermer},
  {Desvignes}, {de Angelis}, {de Palma}, {Digel}, {Dormody}, {Silva}, {Drell},
  {Dubois}, {Dumora}, {Espinoza}, {Farnier}, {Favuzzi}, {Fegan}, {Focke},
  {Frailis}, {Freire}, {Fukazawa}, {Funk}, {Fusco}, {Gargano}, {Gasparrini},
  {Gehrels}, {Germani}, {Giavitto}, {Giglietto}, {Giordano}, {Glanzman},
  {Godfrey}, {Grenier}, {Grondin}, {Grove}, {Guillemot}, {Guiriec}, {Hadasch},
  {Harding}, {Hays}, {Hobbs}, {Horan}, {Hughes}, {J{\'o}hannesson}, {Johnson},
  {Johnson}, {Johnson}, {Johnston}, {Kamae}, {Katagiri}, {Kataoka}, {Kawai},
  {Kerr}, {Kn{\"o}dlseder}, {Kramer}, {Kuss}, {Lande}, {Latronico},
  {Lemoine-Goumard}, {Llena Garde}, {Longo}, {Loparco}, {Lott}, {Lovellette},
  {Lubrano}, {Lyne}, {Makeev}, {Manchester}, {Marelli}, {Mazziotta},
  {McConville}, {McEnery}, {McGlynn}, {Meurer}, {Michelson}, {Mitthumsiri},
  {Mizuno}, {Moiseev}, {Monte}, {Monzani}, {Morselli}, {Moskalenko}, {Murgia},
  {Nolan}, {Norris}, {Noutsos}, {Nuss}, {Ohsugi}, {Omodei}, {Orlando}, {Ormes},
  {Ozaki}, {Paneque}, {Panetta}, {Parent}, {Pelassa}, {Pepe}, {Pesce-Rollins},
  {Pierbattista}, {Piron}, {Porter}, {Rain{\`o}}, {Rando}, {Ransom}, {Razzano},
  {Reimer}, {Reimer}, {Reposeur}, {Ripken}, {Ritz}, {Rochester}, {Rodriguez},
  {Romani}, {Roth}, {Ryde}, {Sadrozinski}, {Sander}, {Saz Parkinson},
  {Scargle}, {Sgr{\`o}}, {Siskind}, {Smith}, {Smith}, {Spandre}, {Spinelli},
  {Stappers}, {Starck}, {Strickman}, {Suson}, {Takahashi}, {Tanaka}, {Thayer},
  {Thayer}, {Theureau}, {Thompson}, {Thorsett}, {Tibaldo}, {Torres}, {Tosti},
  {Tramacere}, {Usher}, {Van Etten}, {Vasileiou}, {Venter}, {Vilchez},
  {Vitale}, {Waite}, {Wallace}, {Wang}, {Weltevrede}, {Winer}, {Wood},
  {Ylinen}, \& {Ziegler}}]{j0034}
---. 2010{\natexlab{a}}, \apj, 712, 957

\bibitem[{{Abdo} {et~al.}(2010{\natexlab{b}}){Abdo}, {Ackermann}, {Ajello},
  {Allafort}, {Antolini}, {Atwood}, {Axelsson}, {Baldini}, {Ballet},
  {Barbiellini}, \& et~al.}]{aaa+2010}
---. 2010{\natexlab{b}}, \apjs, 188, 405

\bibitem[{{Abdo} {et~al.}(2010{\natexlab{c}}){Abdo}, {Ackermann}, {Ajello},
  {Atwood}, {Axelsson}, {Baldini}, {Ballet}, {Barbiellini}, {Baring},
  {Bastieri}, \& et~al.}]{aaa+10}
---. 2010{\natexlab{c}}, \apjs, 187, 460

\bibitem[{{Alpar} {et~al.}(1982){Alpar}, {Cheng}, {Ruderman}, \&
  {Shaham}}]{alp+82}
{Alpar}, M.~A., {Cheng}, A.~F., {Ruderman}, M.~A., \& {Shaham}, J. 1982, \nat,
  300, 728

\bibitem[{{Archibald} {et~al.}(2009)}]{arc+09}
{Archibald}, A.~M., {et~al.} 2009, Science, 324, 1411

\bibitem[{{Atwood} {et~al.}(2009){Atwood}, {Abdo}, {Ackermann}, {Althouse},
  {Anderson}, {Axelsson}, {Baldini}, {Ballet}, {Band}, {Barbiellini}, \&
  et~al.}]{atw+2009}
{Atwood}, W.~B., {Abdo}, A.~A., {Ackermann}, M., {et~al.} 2009, \apj, 697, 1071

\bibitem[{{Bhattacharya} \& {van den Heuvel}(1991)}]{bv91}
{Bhattacharya}, D., \& {van den Heuvel}, E.~P.~J. 1991, \physrep, 203, 1

\bibitem[{{Burderi} {et~al.}(2003){Burderi}, {Di Salvo}, {D'Antona}, {Robba},
  \& {Testa}}]{bur+03}
{Burderi}, L., {Di Salvo}, T., {D'Antona}, F., {Robba}, N.~R., \& {Testa}, V.
  2003, \aap, 404, L43

\bibitem[{{Burgay} {et~al.}(2003)}]{burgay+03}
{Burgay}, M., {et~al.} 2003, \apj, 589, 902

\bibitem[{{Campana} {et~al.}(2004)}]{cam+04}
{Campana}, S., {et~al.} 2004, \apjl, 614, L49

\bibitem[{{Chakrabarty} \& {Morgan}(1998)}]{cm98}
{Chakrabarty}, D., \& {Morgan}, E.~H. 1998, \nat, 394, 346

\bibitem[{{Cheng} {et~al.}(1986){Cheng}, {Ho}, \& {Ruderman}}]{chr86}
{Cheng}, K.~S., {Ho}, C., \& {Ruderman}, M. 1986, \apj, 300, 500

\bibitem[{{Cognard} {et~al.}(2011){Cognard}, {Guillemot}, {Johnson}, {Smith},
  {Venter}, {Harding}, {Wolff}, {Cheung}, {Donato}, {Abdo}, {Ballet}, {Camilo},
  {Desvignes}, {Dumora}, {Ferrara}, {Freire}, {Grove}, {Johnston}, {Keith},
  {Kramer}, {Lyne}, {Michelson}, {Parent}, {Ransom}, {Ray}, {Romani}, {Saz
  Parkinson}, {Stappers}, {Theureau}, {Thompson}, {Weltevrede}, \&
  {Wood}}]{j2017-j2302}
{Cognard}, I., {Guillemot}, L., {Johnson}, T.~J., {et~al.} 2011, \apj, 732, 47

\bibitem[{{D'Avanzo} {et~al.}(2009){D'Avanzo}, {Campana}, {Casares}, {Covino},
  {Israel}, \& {Stella}}]{dav+09}
{D'Avanzo}, P., {Campana}, S., {Casares}, J., {et~al.} 2009, \aap, 508, 297

\bibitem[{{D'Avanzo} {et~al.}(2007){D'Avanzo}, {Campana}, {Covino}, {Israel},
  {Stella}, \& {Andreuzzi}}]{dav+07}
{D'Avanzo}, P., {Campana}, S., {Covino}, S., {et~al.} 2007, \aap, 472, 881

\bibitem[{{Deloye} {et~al.}(2008){Deloye}, {Heinke}, {Taam}, \&
  {Jonker}}]{del+08}
{Deloye}, C.~J., {Heinke}, C.~O., {Taam}, R.~E., \& {Jonker}, P.~G. 2008,
  \mnras, 391, 1619

\bibitem[{{Falanga} {et~al.}(2005){Falanga}, {Kuiper}, {Poutanen}, {Bonning},
  {Hermsen}, {di Salvo}, {Goldoni}, {Goldwurm}, {Shaw}, \& {Stella}}]{fkp+05}
{Falanga}, M., {Kuiper}, L., {Poutanen}, J., {et~al.} 2005, \aap, 444, 15

\bibitem[{{Freire} {et~al.}(2011){Freire}, {Abdo}, {Ajello}, {Allafort},
  {Ballet}, {Barbiellini}, {Bastieri}, {Bechtol}, {Bellazzini}, {Blandford},
  {Bloom}, {Bonamente}, {Borgland}, {Brigida}, {Bruel}, {Buehler}, {Buson},
  {Caliandro}, {Cameron}, {Camilo}, {Caraveo}, {Cecchi}, {{\c C}elik},
  {Charles}, {Chekhtman}, {Cheung}, {Chiang}, {Ciprini}, {Claus}, {Cognard},
  {Cohen-Tanugi}, {Cominsky}, {de Palma}, {Dermer}, {do Couto e Silva},
  {Dormody}, {Drell}, {Dubois}, {Dumora}, {Espinoza}, {Favuzzi}, {Fegan},
  {Ferrara}, {Focke}, {Fortin}, {Fukazawa}, {Fusco}, {Gargano}, {Gasparrini},
  {Gehrels}, {Germani}, {Giglietto}, {Giordano}, {Giroletti}, {Glanzman},
  {Godfrey}, {Grenier}, {Grondin}, {Grove}, {Guillemot}, {Guiriec}, {Hadasch},
  {Harding}, {J{\'o}hannesson}, {Johnson}, {Johnson}, {Johnston}, {Katagiri},
  {Kataoka}, {Keith}, {Kerr}, {Kn{\"o}dlseder}, {Kramer}, {Kuss}, {Lande},
  {Latronico}, {Lee}, {Lemoine-Goumard}, {Longo}, {Loparco}, {Lovellette},
  {Lubrano}, {Lyne}, {Manchester}, {Marelli}, {Mazziotta}, {McEnery},
  {Michelson}, {Mizuno}, {Moiseev}, {Monte}, {Monzani}, {Morselli},
  {Moskalenko}, {Murgia}, {Nakamori}, {Nolan}, {Norris}, {Nuss}, {Ohsugi},
  {Okumura}, {Omodei}, {Orlando}, {Ozaki}, {Paneque}, {Parent},
  {Pesce-Rollins}, {Pierbattista}, {Piron}, {Porter}, {Rain{\`o}}, {Ransom},
  {Ray}, {Reimer}, {Reimer}, {Reposeur}, {Ritz}, {Romani}, {Roth},
  {Sadrozinski}, {Saz Parkinson}, {Shannon}, {Siskind}, {Smith}, {Spinelli},
  {Stappers}, {Suson}, {Takahashi}, {Tanaka}, {Tauris}, {Thayer}, {Theureau},
  {Thompson}, {Thorsett}, {Tibaldo}, {Torres}, {Tosti}, {Troja},
  {Vandenbroucke}, {Van Etten}, {Vasileiou}, {Venter}, {Vianello}, {Vilchez},
  {Vitale}, {Waite}, {Wang}, {Wood}, {Yang}, {Ziegler}, \& {Zimmer}}]{j1823}
{Freire}, P.~C.~C., {Abdo}, A.~A., {Ajello}, M., {et~al.} 2011, Science, 334,
  1107

\bibitem[{{Galloway}(2006)}]{g06}
{Galloway}, D.~K. 2006, in American Institute of Physics Conference Series,
  Vol. 840, The Transient Milky Way: A Perspective for MIRAX, ed. F.~{D'Amico},
  J.~{Braga}, \& R.~E. {Rothschild}, 50--54

\bibitem[{{Galloway} {et~al.}(2002){Galloway}, {Chakrabarty}, {Morgan}, \&
  {Remillard}}]{gcm+02}
{Galloway}, D.~K., {Chakrabarty}, D., {Morgan}, E.~H., \& {Remillard}, R.~A.
  2002, \apjl, 576, L137

\bibitem[{{Guillemot} {et~al.}(2012{\natexlab{a}}){Guillemot}, {Freire},
  {Cognard}, {Johnson}, {Takahashi}, {Kataoka}, {Desvignes}, {Camilo},
  {Ferrara}, {Harding}, {Janssen}, {Keith}, {Kerr}, {Kramer}, {Parent},
  {Ransom}, {Ray}, {Saz Parkinson}, {Smith}, {Stappers}, \& {Theureau}}]{j2043}
{Guillemot}, L., {Freire}, P.~C.~C., {Cognard}, I., {et~al.}
  2012{\natexlab{a}}, \mnras, 422, 1294

\bibitem[{{Guillemot} {et~al.}(2012{\natexlab{b}}){Guillemot}, {Johnson},
  {Venter}, {Kerr}, {Pancrazi}, {Livingstone}, {Janssen}, {Jaroenjittichai},
  {Kramer}, {Cognard}, {Stappers}, {Harding}, {Camilo}, {Espinoza}, {Freire},
  {Gargano}, {Grove}, {Johnston}, {Michelson}, {Noutsos}, {Parent}, {Ransom},
  {Ray}, {Shannon}, {Smith}, {Theureau}, {Thorsett}, \& {Webb}}]{j1959}
{Guillemot}, L., {Johnson}, T.~J., {Venter}, C., {et~al.} 2012{\natexlab{b}},
  \apj, 744, 33

\bibitem[{{Hartman} {et~al.}(2009){Hartman}, {Patruno}, {Chakrabarty},
  {Markwardt}, {Morgan}, {van der Klis}, \& {Wijnands}}]{hpc+09}
{Hartman}, J.~M., {Patruno}, A., {Chakrabarty}, D., {et~al.} 2009, \apj, 702,
  1673

\bibitem[{{Homer} {et~al.}(2001){Homer}, {Charles}, {Chakrabarty}, \& {van
  Zyl}}]{hom+01}
{Homer}, L., {Charles}, P.~A., {Chakrabarty}, D., \& {van Zyl}, L. 2001,
  \mnras, 325, 1471

\bibitem[{{Iacolina} {et~al.}(2009){Iacolina}, {Burgay}, {Burderi}, {Possenti},
  \& {di Salvo}}]{ibb+09}
{Iacolina}, M.~N., {Burgay}, M., {Burderi}, L., {Possenti}, A., \& {di Salvo},
  T. 2009, \aap, 497, 445

\bibitem[{{in 't Zand} {et~al.}(1998){in 't Zand}, {Heise}, {Muller},
  {Bazzano}, {Cocchi}, {Natalucci}, \& {Ubertini}}]{int+98}
{in 't Zand}, J.~J.~M., {Heise}, J., {Muller}, J.~M., {et~al.} 1998, \aap, 331,
  L25

\bibitem[{{Jonker} {et~al.}(2008){Jonker}, {Torres}, \& {Steeghs}}]{jts08}
{Jonker}, P.~G., {Torres}, M.~A.~P., \& {Steeghs}, D. 2008, \apj, 680, 615

\bibitem[{{Keith} {et~al.}(2011){Keith}, {Johnston}, {Ray}, {Ferrara}, {Saz
  Parkinson}, {{\c C}elik}, {Belfiore}, {Donato}, {Cheung}, {Abdo}, {Camilo},
  {Freire}, {Guillemot}, {Harding}, {Kramer}, {Michelson}, {Ransom}, {Romani},
  {Smith}, {Thompson}, {Weltevrede}, \& {Wood}}]{j2241}
{Keith}, M.~J., {Johnston}, S., {Ray}, P.~S., {et~al.} 2011, \mnras, 414, 1292

\bibitem[{{Kerr} {et~al.}(2012){Kerr}, {Camilo}, {Johnson}, {Ferrara},
  {Guillemot}, {Harding}, {Hessels}, {Johnston}, {Keith}, {Kramer}, {Ransom},
  {Ray}, {Reynolds}, {Sarkissian}, \& {Wood}}]{j0101}
{Kerr}, M., {Camilo}, F., {Johnson}, T.~J., {et~al.} 2012, \apjl, 748, L2

\bibitem[{{Larsson}(1996)}]{lar96}
{Larsson}, S. 1996, \aaps, 117, 197

\bibitem[Lewis et 
al.(2010)]{lew10} Lewis, F., Russell, D.~M., Jonker, P.~G., et al.\ 2010, \aap, 517, A72

\bibitem[{{Muslimov} \& {Harding}(2004)}]{mh04}
{Muslimov}, A.~G., \& {Harding}, A.~K. 2004, \apj, 606, 1143

\bibitem[{{Pallanca} {et~al.}(2012){Pallanca}, {Mignani}, {Dalessandro},
  {Ferraro}, {Lanzoni}, {Possenti}, {Burgay}, \& {Sabbi}}]{j0610}
{Pallanca}, C., {Mignani}, R.~P., {Dalessandro}, E., {et~al.} 2012, \apj, 755,
  180

\bibitem[{{Papitto} {et~al.}(2007){Papitto}, {di Salvo}, {Burderi}, {Menna},
  {Lavagetto}, \& {Riggio}}]{pdb+07}
{Papitto}, A., {di Salvo}, T., {Burderi}, L., {et~al.} 2007, \mnras, 375, 971

\bibitem[{{Patruno}(2010)}]{p10}
{Patruno}, A. 2010, \apj, 722, 909

\bibitem[{{Patruno} {et~al.}(2012){Patruno}, {Bult}, {Gopakumar}, {Hartman},
  {Wijnands}, {van der Klis}, \& {Chakrabarty}}]{pbg+12}
{Patruno}, A., {Bult}, P., {Gopakumar}, A., {et~al.} 2012, \apjl, 746, L27

\bibitem[{{Patruno} \& {Watts}(2012)}]{pw12}
{Patruno}, A., \& {Watts}, A.~L. 2012, ArXiv e-prints

\bibitem[{{Ransom} {et~al.}(2011){Ransom}, {Ray}, {Camilo}, {Roberts}, {{\c
  C}elik}, {Wolff}, {Cheung}, {Kerr}, {Pennucci}, {DeCesar}, {Cognard}, {Lyne},
  {Stappers}, {Freire}, {Grove}, {Abdo}, {Desvignes}, {Donato}, {Ferrara},
  {Gehrels}, {Guillemot}, {Gwon}, {Harding}, {Johnston}, {Keith}, {Kramer},
  {Michelson}, {Parent}, {Saz Parkinson}, {Romani}, {Smith}, {Theureau},
  {Thompson}, {Weltevrede}, {Wood}, \& {Ziegler}}]{j0614-j1231-j2214}
{Ransom}, S.~M., {Ray}, P.~S., {Camilo}, F., {et~al.} 2011, \apjl, 727, L16

\bibitem[{{Ruderman} \& {Sutherland}(1975)}]{rs75}
{Ruderman}, M.~A., \& {Sutherland}, P.~G. 1975, \apj, 196, 51

\bibitem[{{Srinivasan}(2010)}]{sri10}
{Srinivasan}, G. 2010, New A Rev, 54, 93

\bibitem[{{Strohmayer} {et~al.}(2003){Strohmayer}, {Markwardt}, {Swank}, \&
  {in't Zand}}]{sms+03}
{Strohmayer}, T.~E., {Markwardt}, C.~B., {Swank}, J.~H., \& {in't Zand}, J.
  2003, \apjl, 596, L67

\bibitem[{{Takata} {et~al.}(2012){Takata}, {Cheng}, \& {Taam}}]{tct12}
{Takata}, J., {Cheng}, K.~S., \& {Taam}, R.~E. 2012, \apj, 745, 100

\bibitem[{{Tam} {et~al.}(2010){Tam}, {Hui}, {Huang}, {Kong}, {Takata}, {Lin},
  {Yang}, {Cheng}, \& {Taam}}]{tam+10}
{Tam}, P.~H.~T., {Hui}, C.~Y., {Huang}, R.~H.~H., {et~al.} 2010, \apjl, 724,
  L207

\bibitem[{{van Paradijs} \& {McClintock}(1995)}]{vm95}
{van Paradijs}, J., \& {McClintock}, J.~E. 1995, {Optical and Ultraviolet
  Observations of X-ray Binaries} (X-ray Binaries, eds.~W.H.G.~Lewin, J.~van
  Paradijs, and E.P.J.~van den Heuvel, Cambridge: Cambridge Univ.~Press), 58

\bibitem[{{Wang} {et~al.}(2009{\natexlab{a}}){Wang}, {Archibald},
  {Thorstensen}, {Kaspi}, {Lorimer}, {Stairs}, \& {Ransom}}]{wat+09}
{Wang}, Z., {Archibald}, A.~M., {Thorstensen}, J.~R., {et~al.}
  2009{\natexlab{a}}, \apj, 703, 2017

\bibitem[{{Wang} {et~al.}(2009{\natexlab{b}}){Wang}, {Bassa}, {Cumming}, \&
  {Kaspi}}]{wan+09}
{Wang}, Z., {Bassa}, C., {Cumming}, A., \& {Kaspi}, V.~M. 2009{\natexlab{b}},
  \apj, 694, 1115

\bibitem[Wang et al.(2013)]{wan+13} Wang, Z., Breton, R.~P., 
Heinke, C.~O., Deloye, C.~J., \& Zhong, J.\ 2013, \apj, 765, 151

\bibitem[Wang 
\& Hirotani(2011)]{wang11} Wang, R.-B., \& Hirotani, K.\ 2011, \apj, 736, 127

\bibitem[{{Wang} {et~al.}(2012){Wang}, {Wang}, \& {Morrell}}]{wwm12}
{Wang}, X., {Wang}, Z., \& {Morrell}, N. 2012, ArXiv e-prints

\bibitem[{{Wijnands} \& {van der Klis}(1998)}]{wv98}
{Wijnands}, R., \& {van der Klis}, M. 1998, \nat, 394, 344

\end{thebibliography}

\clearpage
\begin{deluxetable}{lcccccccc}
\tabletypesize{\scriptsize}
\tablecaption{Observational properties of the four AMP targets}
\tablewidth{0pt}
\startdata
\hline
\hline
Source name & $d$ & $\nu$ & $\dot{\nu}$ & $\dot{E_{sd}}$ & $\tau$ & Quiescence & $L_{\gamma}$ & $\eta_{\gamma}$ \\
 & (kpc) & ($\mathrm{Hz}$) & ($10^{-15}$ Hz s$^{-1}$) & ($10^{33}$ ergs s$^{-1}$) & (10$^{9}$ yrs) & (MJD) & ($10^{33}$ ergs s$^{-1}$) &  \\
\hline
SAX J1808.4$-$3658 & 3.5$^{a}$ & 400.97521024$^{f}$ & -0.55$^{f}$ & 9 & 11.56 & 54682$-$54720 & $<$5.1 & $<$0.57 \\
 &  &  &  &  &  & 54780$-$55855 &  &  \\
 &  &  &  &  &  & 55915$-$56116 &  &  \\
IGR J00291$+$5934 & 5$^{b}$ & 598.89213061$^{g}$ & -3.0$^{g}$ & 70 & 3.17 & 54777$-$56116 & $<$2.1 & $<$0.03 \\
XTE J1814$-$338 & 8$^{c}$ & 314.35610879$^{h}$ & -- & -- & -- & -- & $<$12 & -- \\
XTE J0929$-$314 & 6$^{d,e}$ & 185.105259$^{i}$ & -- & -- & -- & -- & $<$2.2 & -- \\
\enddata
\tablecomments{Column 5 are the spin-down luminosities of the sources with 
spin-down rate measurements. Column 6 lists the characteristic ages 
derived by $\nu/2\dot{\nu}$. Column 7 lists the epoch of quiescent state defined in this paper. Column 8 lists the 100 MeV to 300 
GeV $\gamma$-ray luminosity upper limits resulting from binned likelihood 
analysis of the LAT data with $\Gamma$ and $E_{cut}$ of 
the spectral models of these sources fixed to 1.4 and 1.6 GeV. Column 9 are
the 100 MeV to 300 GeV $\gamma$-ray conversion efficiency upper limits for the
sources with measurements of spin-down rates. a: \citealt{g06}. 
b: \citealt{fkp+05}. c: \citealt{sms+03}. d: \citealt{gcm+02}. 
e: \citealt{ibb+09}. f: The epoch time is MJD 52499.9602472 
\citep{hpc+09}. g: The epoch time is MJD 54692.0 \citep{p10}. 
h: The epoch time is MJD 52797.27387859868 \citep{pdb+07}. 
i: The epoch time is MJD 52405.48676 \citep{ibb+09}.}
\end{deluxetable}
\normalsize
\noindent

\begin{deluxetable}{lccccccccc}
\tabletypesize{\scriptsize}
\tablecaption{Observational properties of the 20 confirmed $\gamma$-ray MSPs in the \textit{Fermi} 2-year catalog.}
\tablewidth{0pt}
\startdata
\hline
\hline
Source Name & d & $\nu$ & $\dot{\nu}$ & $\dot{E_{sd}}$ & $\tau$ & $L_{\gamma}$ & $\eta_{\gamma}$ \\
 & (kpc) & (Hz) & ($10^{-15}$ Hz s$^{-1}$) & ($10^{33}$ ergs s$^{-1}$) & (10$^{9}$ yrs) & ($10^{33}$ ergs s$^{-1}$) & \\
\hline
J0030$+$0451$^{\ast}$ & 0.3$^{a}$ & 205.3$^{b}$ & -0.4$^{b}$ & 3 & 7.72 & $\mathrm{0.67\pm0.02}$ & $\mathrm{0.197\pm0.007}$ \\
J0034$-$0534 & 0.5$^{c}$ & 531.9$^{c}$ & -1.4$^{c}$ & 30 & 5.96 & $\mathrm{0.55\pm0.05}$ & $\mathrm{0.018\pm0.002}$ \\
J0101$-$6422 & 0.6$^{d}$ & 389.1$^{d}$ & -0.8$^{d}$ & 12 & 7.84 & $\mathrm{0.39\pm0.04}$ & $\mathrm{0.032\pm0.004}$ \\
J0218$+$4232$^{\ast}$ & 2.7$^{a}$ & 431.0$^{b}$ & -14.3$^{b}$ & 243 & 0.48 & $\mathrm{41\pm2}$ & $\mathrm{0.17\pm0.01}$ \\
J0437$-$4715$^{\ast}$ & 0.2$^{a}$ & 173.6$^{b}$ & -0.4$^{b}$ & 3 & 6.52 & $\mathrm{0.054\pm0.004}$ & $\mathrm{0.018\pm0.001}$ \\
J0610$-$2100 & 3.5$^{e}$ & 263.2$^{e}$ & -0.9$^{e}$ & 9 & 4.88 & $\mathrm{12\pm2}$ & $\mathrm{1.3\pm0.3}$ \\
J0613$-$0200$^{\ast}$ & 0.5$^{a}$ & 326.8$^{b}$ & -1.0$^{b}$ & 12 & 5.39 & $\mathrm{0.85\pm0.06}$ & $\mathrm{0.068\pm0.005}$ \\
J0614$-$3329 & 1.9$^{f}$ & 317.5$^{f}$ & -1.8$^{f}$ & 22 & 2.85 & $\mathrm{48\pm1}$ & $\mathrm{2.2\pm0.1}$ \\
J0751$+$1807$^{\ast}$ & 0.6$^{a}$ & 287.4$^{b}$ & -0.5$^{b}$ & 6 & 9.20 & $\mathrm{0.67\pm0.07}$ & $\mathrm{0.12\pm0.01}$ \\
J1231$-$1411 & 0.4$^{f}$ & 271.7$^{f}$ & -1.7$^{f}$ & 18 & 2.56 & $\mathrm{2.02\pm0.06}$ & $\mathrm{0.112\pm0.003}$ \\
J1614$-$2230$^{\ast}$ & 1.3$^{a}$ & 317.5$^{b}$ & -0.4$^{b}$ & 5 & 12.49 & $\mathrm{4.6\pm0.4}$ & $\mathrm{0.91\pm0.08}$ \\
J1744$-$1134$^{\ast}$ & 0.5$^{a}$ & 245.7$^{b}$ & -0.4$^{b}$ & 4 & 9.22 & $\mathrm{0.94\pm0.07}$ & $\mathrm{0.23\pm0.02}$ \\
J1823$-$3021A & 8.4$^{g}$ & 183.8$^{g}$ & -114.2$^{g}$ & 829 & 0.03 & $\mathrm{133\pm23}$ & $\mathrm{0.16\pm0.03}$ \\
J1959$+$2048 & 2.5$^{h}$ & 625.0$^{h}$ & -6.6$^{h}$ & 163 & 1.50 & $\mathrm{12\pm1}$ & $\mathrm{0.077\pm0.009}$ \\
J2017$+$0603 & 1.6$^{i}$ & 344.8$^{i}$ & -1.0$^{i}$ & 13 & 5.54 & $\mathrm{11.0\pm0.7}$ & $\mathrm{0.82\pm0.05}$ \\
J2043$+$1710 & 1.8$^{j}$ & 420.2$^{j}$ & -0.9$^{j}$ & 15 & 7.20 & $\mathrm{11.4\pm0.8}$ & $\mathrm{0.74\pm0.05}$ \\
J2124$-$3358$^{\ast}$ & 0.3$^{a}$ & 202.8$^{b}$ & -0.5$^{b}$ & 4 & 6.51 & $\mathrm{0.28\pm0.01}$ & $\mathrm{0.070\pm0.004}$ \\
J2214$+$3000 & 1.5$^{f}$ & 320.5$^{f}$ & -1.4$^{f}$ & 18 & 3.53 & $\mathrm{8.9\pm0.5}$ & $\mathrm{0.49\pm0.03}$ \\
J2241$-$5236 & 0.5$^{k}$ & 457.3$^{k}$ & -1.4$^{k}$ & 25 & 5.19 & $\mathrm{1.02\pm0.06}$ & $\mathrm{0.042\pm0.002}$ \\
J2302$+$4442 & 1.2$^{i}$ & 192.7$^{i}$ & -0.5$^{i}$ & 4 & 6.19 & $\mathrm{6.7\pm0.4}$ & $\mathrm{1.8\pm0.1}$ \\
\enddata
\tablecomments{Observational properties of the 20 confirmed $\gamma$-ray MSPs in Fermi 2-year catalog. The sources marked with $\ast$ are the first eight $\gamma$-ray MSPs detected by Fermi/LAT. Column 2 lists the distance of each source. Columns 3 and 4 list the frequency and spin-down rate of each source. Column 5 lists the spin-down luminosities of sources. Column 6 lists the characteristic ages derived by $\nu/2\dot{\nu}$. Column 7 lists the 100 MeV $-$ 100 GeV $\gamma$-ray luminosities derived from catalog. Column 8 lists the 100 MeV $-$ 100 GeV $\gamma$-ray conversion efficiencies. The uncertainties of luminosities and efficiencies are derived from flux uncertainties, the distance uncertainties are not considered here. a: \citealt{aaa+09}. b: \citealt{aaa+10}. c: \citealt{j0034}. d: \citealt{j0101}. e: \citealt{j0610}. f: \citealt{j0614-j1231-j2214}. g: \citealt{j1823}. h: \citealt{j1959}. i: \citealt{j2017-j2302}. j: \citealt{j2043}. k: \citealt{j2241}.}
\end{deluxetable}
\normalsize
\noindent

\newpage

\begin{figure}
\begin{center}
\epsscale{1.0}
\plottwo{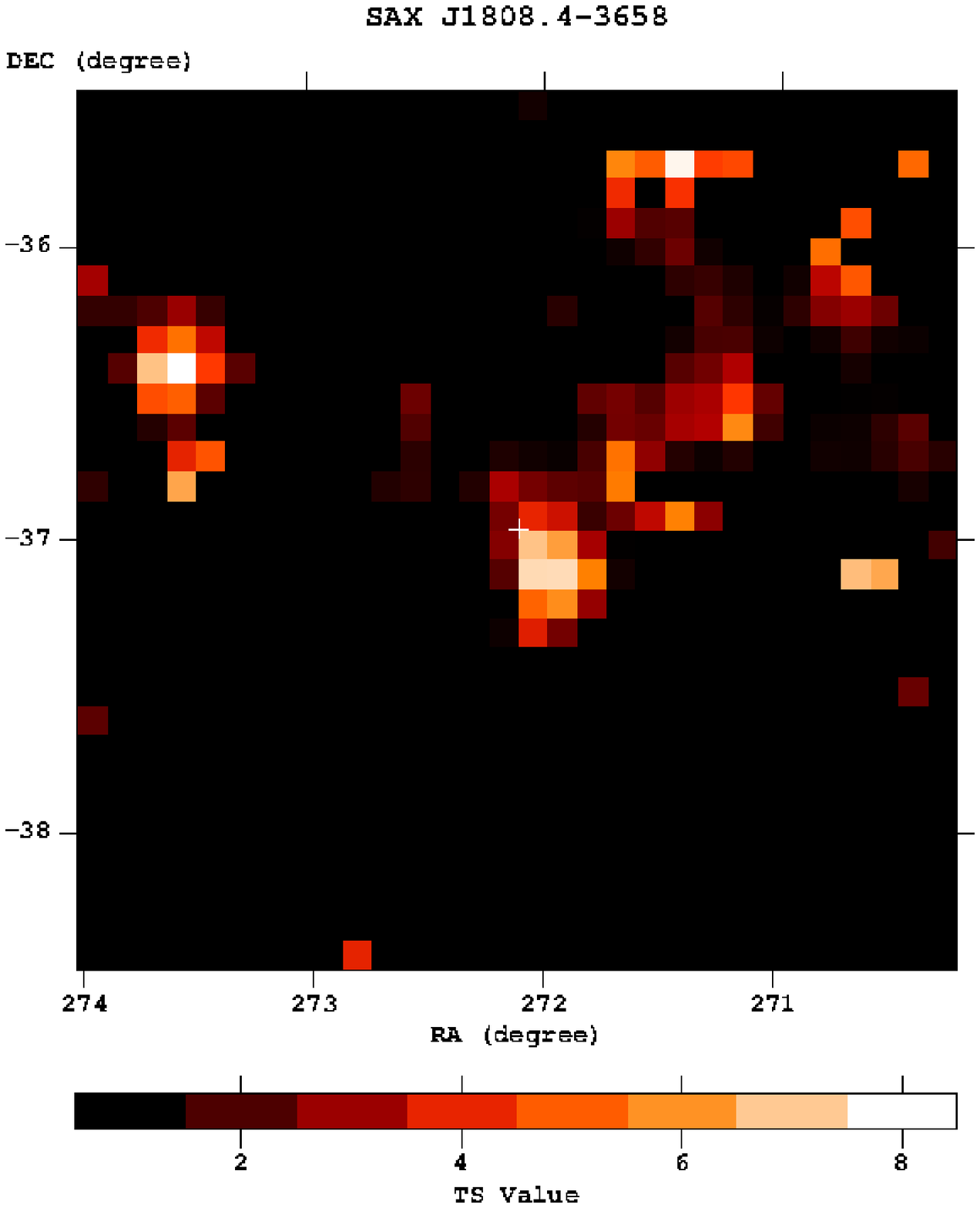}{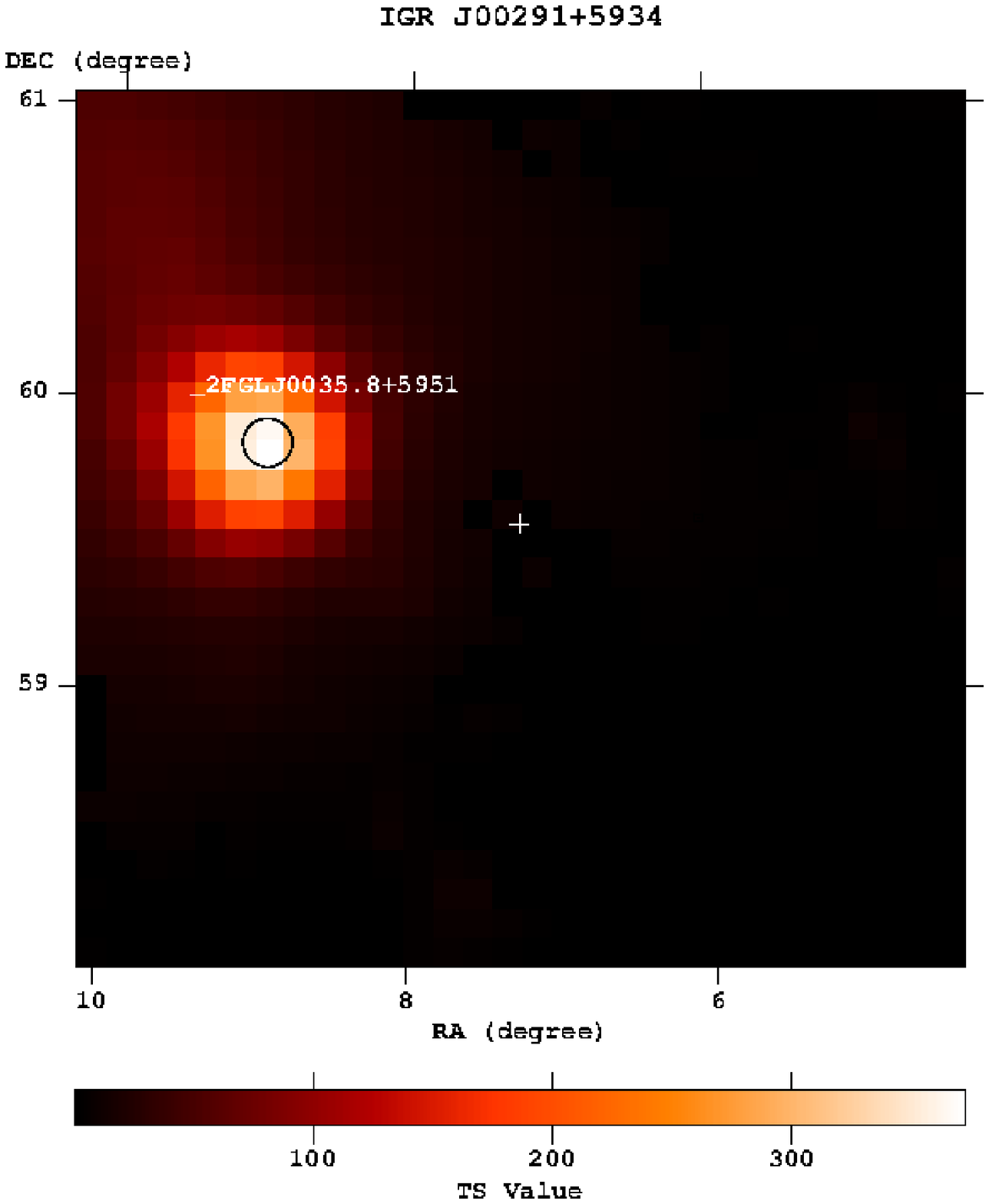}
\plottwo{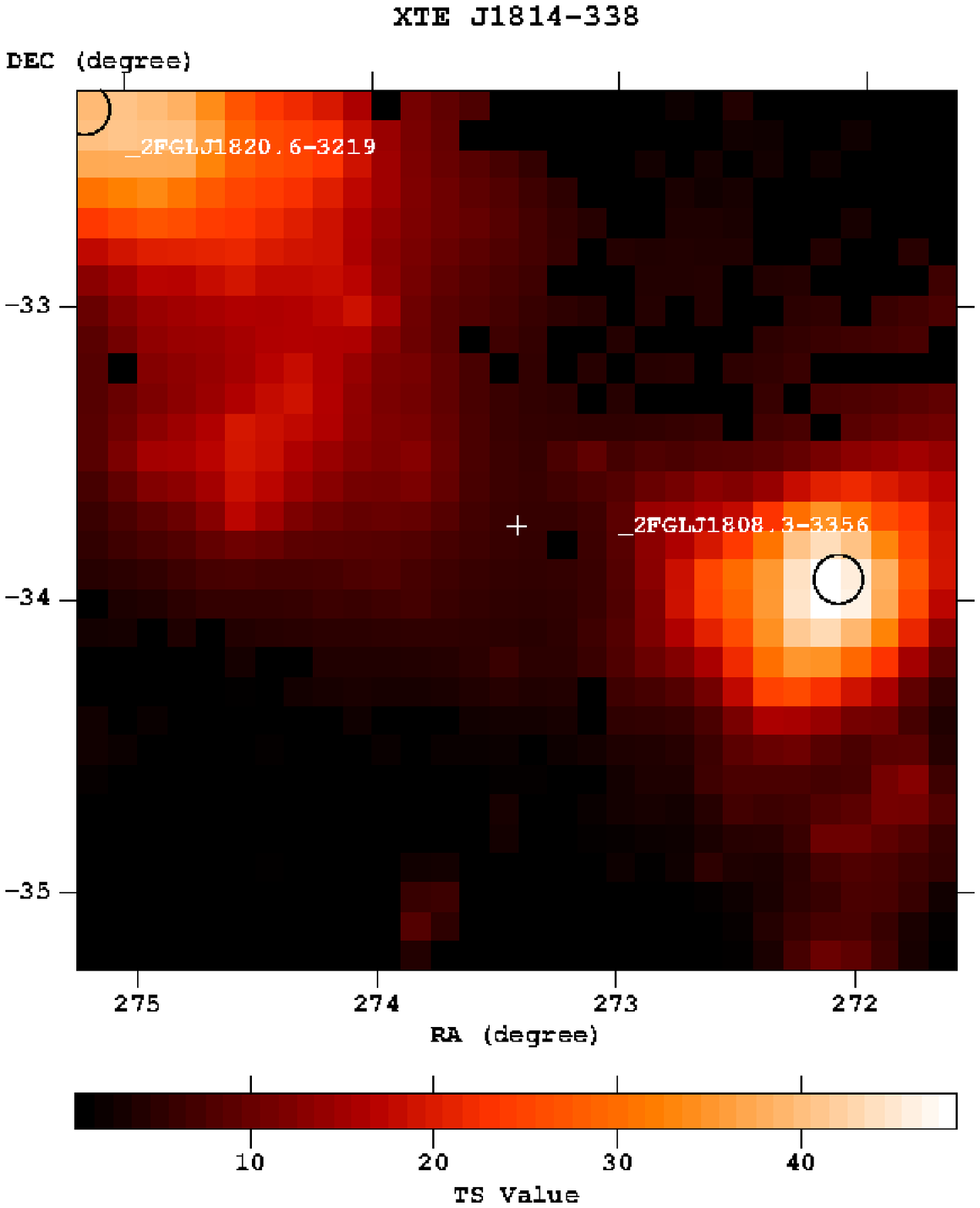}{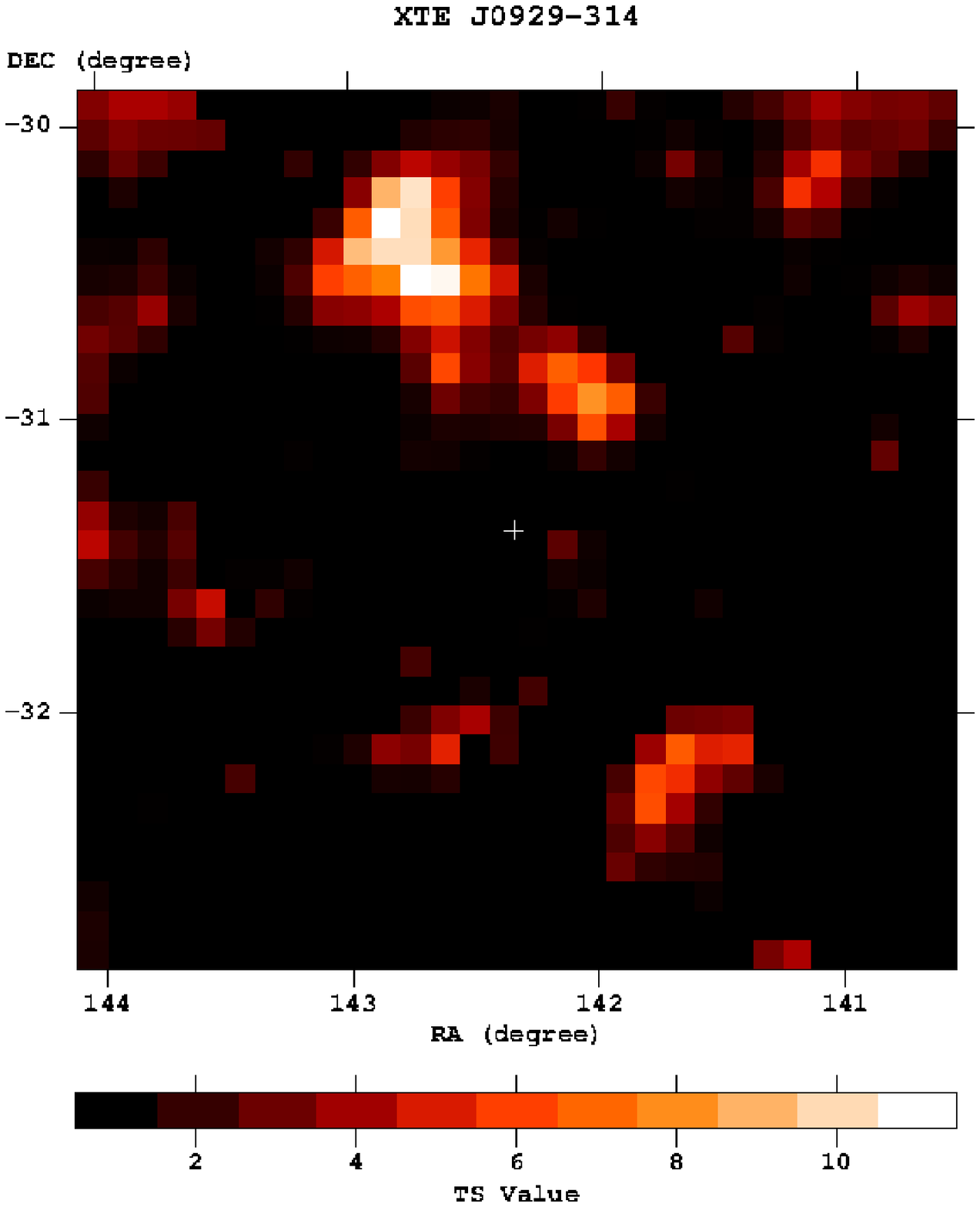}
\caption{200 MeV $-$ 300 GeV TS maps of $\mathrm{3^{o}\times3^{o}}$ regions centered at the four AMPs. The image scales of the maps are 0\arcdeg.1 pixel$^{-1}$ and the crosses in the center of each maps mark the positions of the AMPs. The circles in two of these maps mark the $\gamma$-ray sources reported in the Fermi 2-year catalog.}
\end{center}
\end{figure}

\begin{figure}
\epsscale{1.0}
\plotone{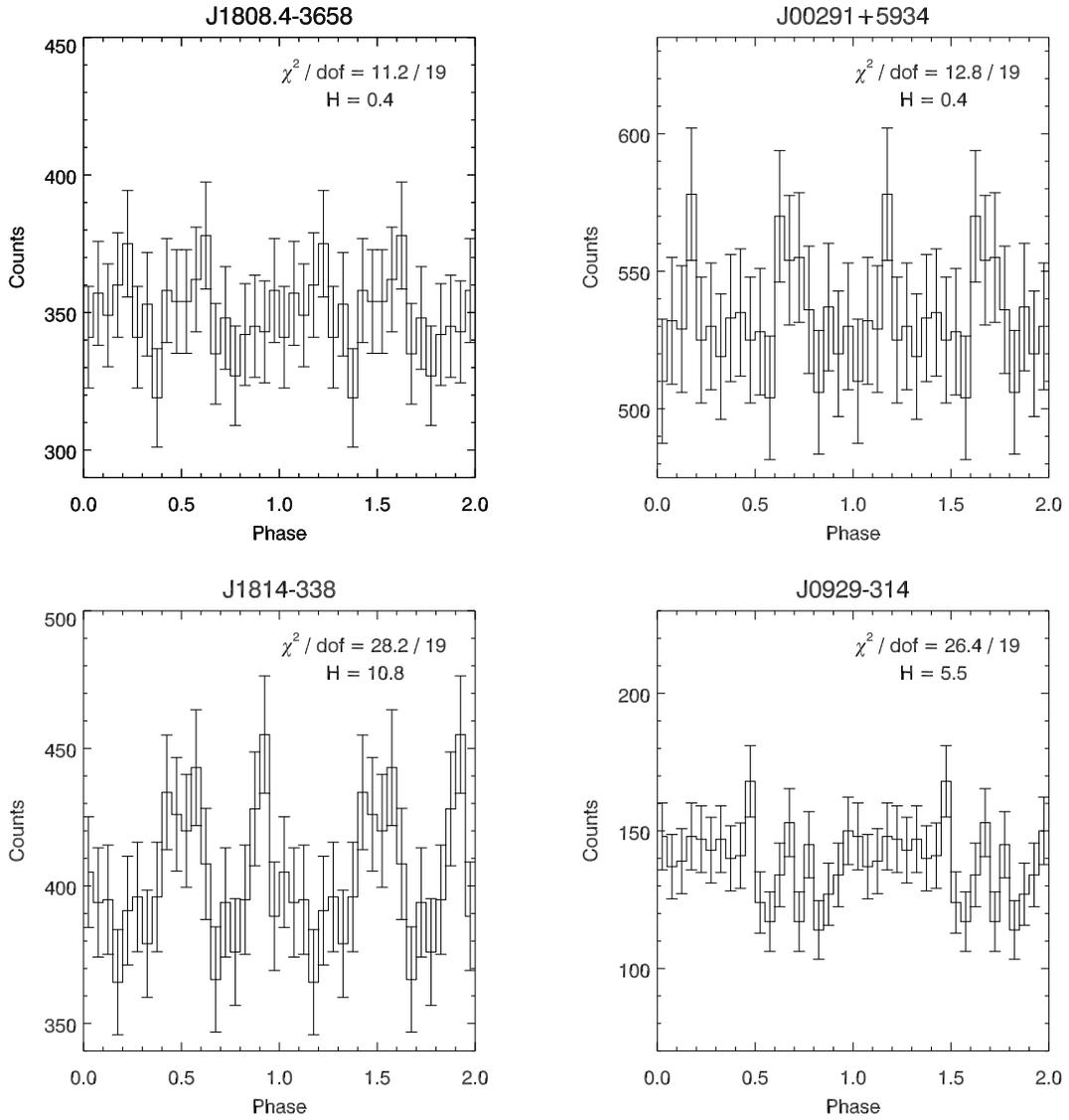}
\caption{100 MeV $-$ 300 GeV folded light curves of the four AMPs included in our work. The phase resolution of each light curve is 0.05. The spin parameters we used in epoch folding are those listed in Table 1. 
For each light curve we made the $\chi^{2}$ test and H test to identify possible presence of pulsations. The obtained $\chi^{2}$/dof 
and H values are given in the upper right corner of each panel.}
\end{figure}

\begin{figure}
\epsscale{1.0}
\plotone{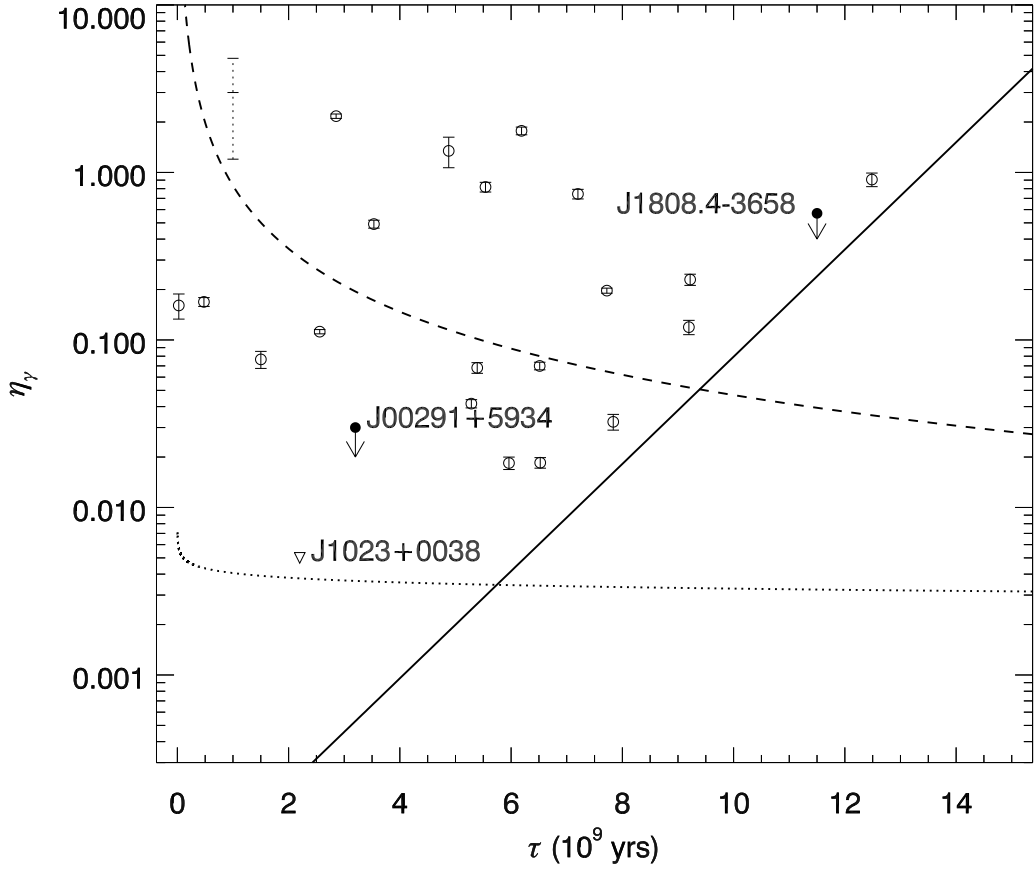}
\caption{$\gamma$-ray conversion efficiencies and characteristic ages of MSPs,
with our two AMP targets included. 
The open circles represent the confirmed MSPs in the \textit{Fermi} 2-year 
catalog, and the filled circles with arrows represent SAX J1808.4$-$3658 
and IGR J00291$+$5934. The open triangle represents the newly born MSP 
PSR J1023$+$0038. The errors of data points are derived from flux 
uncertainties. The dotted bar in the top left of the figure represents errors 
derived from distance uncertainties which are assumed to be 30$\%$ of 
the distance values. A solid line is defined as $\eta_{\gamma}=10^{0.32\tau_9-4.3}$ to suggest that currently older detectable MSPs have higher $\eta_{\gamma}$. For a comparison, the death line given by
\citet{wang11} is plotted as the dotted curve ($L_{\gamma}=2\times10^{30}$ ergs s$^{-1}$
is used) and the relation $L_{\gamma}\sim10^{34}\tau_9^{-5/4}$ ergs s$^{-1}$ given by \citet{tct12} is plotted
as the dashed curve (an approximate average value of $\dot{E_{sd}}=1.2\times10^{34}$ ergs s$^{-1}$ from the confirmed $\gamma$-ray MSPs is used; Table 2).}
\end{figure}

\end{document}